\documentclass[12pt,nofootinbib,preprint]{revtex4}
\usepackage{graphicx}
\usepackage{color}
\newcommand{\be}{\begin{equation}}
\newcommand{\ee}{\end{equation}}
\newcommand{\br}{\begin{eqnarray}}
\newcommand{\er}{\end{eqnarray}}

\def\ptm{E{\!\!\!/}_T} 
\def\beq{\begin{equation}}
\def\eeq{\end{equation}}
\def\bnq{\begin{eqnarray}}
\def\enq{\end{eqnarray}}
\def\barr{\begin{array}}
\def\earr{\end{array}}
\def\dis{\displaystyle}
\def\lsim{\buildrel{\scriptscriptstyle <}\over{\scriptscriptstyle\sim}}
\def\gsim{\buildrel{\scriptscriptstyle >}\over{\scriptscriptstyle\sim}}
\def\gev{\, {\rm GeV}}
\def\lapp{\mathrel{\rlap{\raise.5ex\hbox{$<$}}
                    {\lower.5ex\hbox{$\sim$}}}}
\def\gapp{\mathrel{\rlap{\raise.5ex\hbox{$>$}}
                    {\lower.5ex\hbox{$\sim$}}}}
\begin{document}
\bibliographystyle{revtex}

\title{LHC signals of $T$-odd heavy quarks in the Littlest Higgs model}
\author{Debajyoti Choudhury\footnote{debchou@physics.du.ac.in}
and Dilip Kumar Ghosh\footnote{dkghosh@physics.du.ac.in}
}
\affiliation{\vspace*{0.1in}Department of Physics $\& $ Astrophysics, \\ 
University of Delhi, Delhi-110007, India}

\vspace*{1.in}
\begin{abstract}
Recently proposed Little Higgs models present a viable solution to
the naturalness problem of the Standard Model. An additional discrete
symmetry, called $T$-parity, has been included in the simplest Little
Higgs models to evade the constraints arising from electroweak
precision data. The Littlest Higgs model with $T$-parity (LHT) not only
predicts a set of new fermions in addition to the heavy gauge bosons
of the original Little Higgs model, but also provides a new candidate
for dark matter. In this paper, we study two particularly 
interesting signatures of $T$-odd fermion pair
production at the LHC, namely,
 $(a)$~$ jj + \ell^+\ell^- +\ptm $ and 
 $(b)$~$ jj + b \bar b + \ell^\pm +\ptm $.
Using a parton level Monte Carlo event
generator, we evaluate both the signal as well as the standard model
background profile for a selected set of model parameters thereby 
developing a good discriminator. Finally, 
we scan the  parameter space and delineate 
the possible discovery region in the same.
\end{abstract}

\maketitle

\section{Introduction}
\label{sect:Intro}
The experimental observation of the Higgs boson(s) and the
determination of its (their) properties is crucial for the
understanding of Electro-Weak Symmetry Breaking (EWSB)
and hence constitutes one the major goals of the presently
operating high energy collider viz. the Tevatron (Run II) as well as future ones
such as the forthcoming LHC and the planned International Linear
Collider (ILC). This process is rendered even more complicated 
by the fact that within the Standard Model (SM), the Higgs boson 
mass is not predicted uniquely. 
Negative results from current search efforts, thus, serve only to set a
lower bound of $114$ GeV on its mass\cite{Eidelman:2004wy,lep}.  
Precision electroweak data, on the other hand,
favor a light Higgs boson with a mass $m_H \leq 186 \,$ GeV at 95\%
CL \cite{higmax}.

This immediately leads us to the fine-tuning problem in the SM, namely 
that there is no symmetry which can protect the Higgs mass 
$M_h$ from large radiative corrections from the ultra-violet. 
As this constitutes an outstanding theoretical problem with the SM,
several mechanisms to protect the Higgs mass have been 
proposed; examples include technicolor, supersymmetry and 
a low fundamental quantum gravity scale.
Of these, supersymmetry is especially popular
as the stabilization of $M_h$ is assured in a {\em natural manner} due 
to the symmetry between the bosonic and fermionic degrees of freedom in 
the theory. On the other hand, technicolor theories solve the
hierarchy problem by introducing some strong interactions at scales not too 
much above the electroweak scale. The low scale fundamental quantum gravity
models resolve the issue by just lowering the fundamental Planck scale.
Unfortunately though, 
despite intensive efforts over decades, no experimental hint 
for any of these scenarios has been forthcoming. 
Consequently, it is very
important to explore alternative mechanisms for EWSB that are 
testable in current or forthcoming experiments.
Recently, such an alternative mechanism for solving the naturalness problem
of the standard model has been developed~\cite{lh0}.
Dubbed as Little Higgs models, these incorporate the SM
Higgs as a pseudo-Goldstone boson of some global symmetry which is 
spontaneously broken at a high scale $\Lambda ( \equiv 4\pi f) \sim 10 $ TeV. 
The low energy 
effective theory is described by a non-linear sigma model.
With the introduction of new gauge bosons and partners of the 
top quark with masses of 
the order of $f$, the quadratically divergent contributions to
the Higgs mass are exactly cancelled at the one loop level, thereby 
ameliorating the fine-tuning problem.

However, in the presence of such a plethora of new particles, the
electroweak observables receive additional contributions at the tree
level due to the exchange of heavy gauge bosons (as also from a
non-zero vacuum expectation of a triplet Higgs field that often 
comes about naturally).  These additional contributions are in direct conflict
with experimental data, unless the scale $f$ is above $\sim 5$
TeV\cite{lh_ew}. For such a large value of $f$, one faces the
re-introduction of a fine
tuning between the cutoff scale ($\sim 4\pi f$) for the model 
and the weak scale.  To circumvent
this serious problem of the original Little Higgs model, a new
discrete symmetry, called $T$-parity (and analogous to the $R$ parity in
the MSSM), has been introduced. The Littlest
Higgs Model with $T$-parity (LHT) \cite{lht1,lht2,lht3,lht4} provides
a fully realistic and consistent model which satisfies the electroweak
precision data. Under this new symmetry all standard model fields are
$T$-even, while the new heavy partners are $T$-odd. As a consequence,
all $T$-odd fields can only be generated in pairs. Furthermore, after
the electroweak symmetry breaking, mixing between standard model
gauge bosons with their $T$-odd counterparts is prohibited by this
new discrete symmetry.  Hence, there are no tree level contributions
from $T$-odd heavy partners of the standard model particles to the
electroweak precision observables. With all such corrections arising
only at the one loop level or beyond, these are naturally small.  As
a result of this, the electroweak precision data now allows for 
a relatively low
value of new particle mass scale $f\sim 500$ GeV \cite{lht3}, thereby 
leading to copious production of different $T$-odd heavy
partners of the standard model particles at the LHC as well as future
$e^+ e^-$ linear collider (ILC)~\cite{lht2,wyler,belyaev,carena,kingman}.
Another interesting feature of $T$-parity is the existence of a
neutral and colorless weakly interacting stable $T$-odd particle (LTP)
$A_H$, the heavy partner of the hypercharge gauge boson; very often
termed the {\it heavy photon}, it is a good candidate for cold
dark matter \cite{lht_dm}. 

The long waited $pp$ Large Hadron Collider (LHC), to be operative in a year
from now, will be of great importance in revealing
the mystery of the electroweak symmetry breaking. While the major thrust
would be on the discovery of the standard model Higgs, it will also
provide a great opportunity to explore alternate mechanisms of the
electroweak symmetry breaking. This has motivated some phenomenological
studies of the Littlest Higgs model with $T$-parity
\cite{wyler,belyaev,carena,kingman,biswarup}. In this paper, we revisit the
LHC signatures of the first two generation $T$-odd heavy quark
pair production within the Littlest Higgs model(LHT) \cite{belyaev, wyler,
carena}. Performing a detailed estimation of the
observability of two type of signals $(a)$ $jj + \ell^+\ell^- + \ptm$
and $(b)$ $jj + b {\bar b} +\ell^\pm + \ptm$, we provide the discovery region
at the LHC of the LHT parameter space. The
rest of the paper is organized as follows. In Section
\ref{sect:model}, we briefly discuss the main features of the
model. In Section \ref{sect:t-oddprod}, we discuss pair production of
$T$-odd heavy quarks and its two body decay branching ratio into
standard model quarks and $T$-odd heavy gauge bosons. In Section
\ref{sect:sigback}, signal and background events are
discussed in detail. In section \ref{sect:discover}, we discuss the possible 
$5\sigma $ discovery region in the LHT parameter space using the
signal $(b)$. Finally, our conclusions are given in Section
\ref{sect:concls}.

\section{The Model}
\label{sect:model}
The Littlest Higgs model with $T$-parity has been studied in great
detail elsewhere \cite{lht1,lht2,lht3}, and here we briefly discuss
some important features of the model relevant for our analysis.  It is
a non-linear sigma model based on a $SU(5)$ global symmetry of which a
$[SU(2)_1\times U(1)_1] \times [SU(2)_2\times U(1)_2] $ subgroup is gauged.  
A discrete symmetry ($T$-parity), 
exchanging the two $[SU(2) \times U(1)]$ groups is naturally
introduced in the model.  At a scale $f$, the global symmetry is
spontaneously broken down to a $SO(5)$ group resulting in 14 massless
Nambu-Goldstone (NG) bosons \cite{lh0}.  Simultaneously, the gauged
symmetry is broken down to its subgroup $SU(2)_L\times U(1)_{\rm Y} $
identified as the standard model gauge group.  Consequently, of the 14
NG bosons, four are eaten by the heavy gauge bosons associated with
the broken symmetry. The remaining NG bosons decompose into a $T$-even
$SU(2)$ doublet $h$, considered to be the standard model Higgs
doublet, and a complex 
$T$-odd $SU(2)$ triplet $\Phi $, which acquires a mass
$M_{\Phi} = \sqrt{2} M_{h} f/v_{\rm SM}$ at one loop, with $M_h$ being
the standard model Higgs mass.  These Higgs bosons remain in the low
energy effective theory.

After electroweak symmetry breaking, the masses of the $T$-odd heavy 
partners of the photon $(A_H)$, $Z$-boson $(Z_H)$ and $W$-boson$(W_H)$ are 
given by 
\beq
\barr{rclcl}
& & M_{A_H} & \simeq &  \dis 
\frac{g^\prime f}{\sqrt{5}}\left[1 - \frac{5v^2_{\rm SM}}{8 f^2}
+...\right] \ ; 
\\[2ex]
M_{Z_H} & \simeq &  M_{W_H} & = & \dis g f \left[ 1 - \frac{5v^2_{\rm SM}}{8 f^2}+...\right] \ .
   \label{eq:gauge_mass}
\earr
\eeq
Here, $v_{\rm SM}\simeq 246 $ GeV is the electroweak symmetry breaking scale.
Since $g^\prime < g $,  $A_H$ is substantially lighter 
than other $T$-odd heavy gauge bosons. 

For consistent implementation of $T$-parity in the fermion sector, 
each  standard model fermion doublet must be replaced by a pair of 
fields $F_\alpha (\alpha =1,2)$ \cite{lht1,lht2,lht3}, where each 
$F_\alpha$ is a doublet under $SU(2)_\alpha$ and singlet under the other. 
The aforementioned $T$-parity
exchanges $F_1 $ and $F_2$. The $T$-even combination of $F_\alpha$ is 
identified with 
the standard model fermion doublet and the other ($T$-odd) combination
is its heavy partner $(F_H)$. To generate mass terms for these $T$-odd
heavy fermions through Yukawa interactions one requires additional $T$-odd 
$SU(2)$ singlet fermions in the theory as suggested in \cite{lht1,lht2,lht3}. 
Assuming universal and flavour diagonal Yukawa coupling $\kappa $, 
we have, for $U_H $ and $D_H$ (the $T$-odd heavy partners of 
the standard model quarks $(u,c)$ and $(d,s)$ respectively),  
\bnq
M_{D_H} \simeq \sqrt{2} \, \kappa \, f \ ,
\qquad
M_{U_H}  \simeq \sqrt{2} \, \kappa \, f \, 
         \left(1 - \, \frac{v^2_{\rm SM}}{8 \, f^2} \right) \ .
\label{eq:Toddmass}
\enq 
Since $f \gapp 500 \gev$, 
it is clear from eq.(\ref{eq:Toddmass}) that the up and down 
type $T$-odd heavy partners have nearly equal masses.  We will not discuss the
top sector of the model, since in this paper our main focus will be on the
first two generation heavy quarks. Further details
about the implementation of $T$-parity in the fermion sector including the
top quark sector can be found in Refs.\cite{lht1,lht2,lht3,belyaev}. 
In summary, the complete spectrum of the 
Littlest Higgs model(LHT) with $T$-parity relevant for our analysis will only
depend on two free parameters: the new physics scale $f$ and the flavour 
independent Yukawa coupling $\kappa $ whose range is 
$0.5 \leq \kappa \leq 1.5 $ \cite{lht2,lht3}.

\section{The $1^{\rm st}$ and $2^{\rm nd}$ generation $T$-odd heavy quark 
production and decay}
\label{sect:t-oddprod}
Based on the model of section \ref{sect:model}, we now calculate the leading 
order production rates of $T$-odd quarks at the LHC. 
The latter can be copiously pair produced ($Q_H \bar Q_H$) as long as their 
masses are not too large. With the dominant production
mechanism being the QCD one (both $q \bar q$ and $g g $ initiated), 
one may safely neglect the sub-dominant weak production amplitudes. 
In fact, the latter contributions to $Q_H \bar Q_H $ 
production are even smaller 
than those leading 
to electroweak processes such as $u u \to U U$ or $ d d \to D D$. Although 
the last-mentioned lead to interesting final states containing like-sign 
dilepton pairs, we choose to neglect these.in the current analysis.

\begin{figure}[!h]
\begin{center}
\includegraphics[scale=0.8]{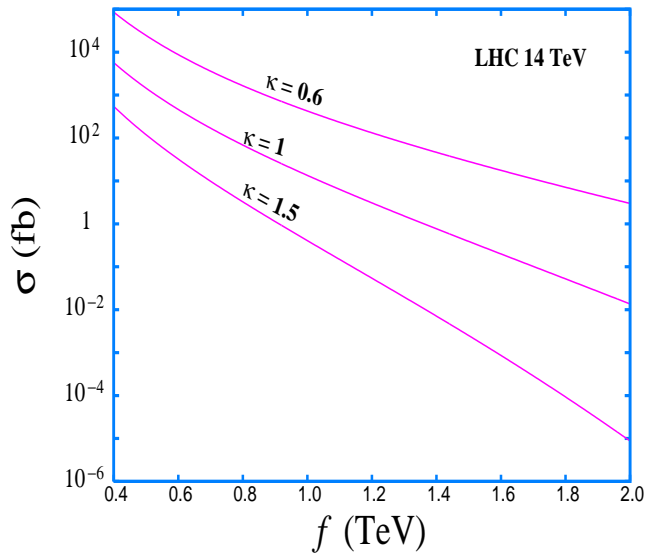}
\end{center}
\caption{\em The variation of the leading order $T$-odd quark pair $(
U_H \bar U_H+D_H \bar D_H+C_H \bar C_H+S_H \bar S_H)$ production 
with the scale $f$ for $\kappa = 0.6,1 $ and $1.5$.}
\label{fig:csprod}
\end{figure}

As the heavy quarks corresponding to the first two generations are
nearly degenerate, and lead to very
similar final state configurations, we sum over all four flavours.  In
our numerical analysis, we use the CTEQ5L parton distribution
functions\cite{cteq5}.  Variation of the factorisation scale over the
range $m_{Q_H}^2 / 4 < Q^2 < 4 m_{Q_H}^2$ corroborates the naive expectation
of the signal cross-section falling off with an increase in the scale,
and, to be conservative, we choose $Q^2 = 4 m_{Q_H}^2$.  In
Fig.~\ref{fig:csprod}, we display the production rate of the $T$-odd
quark as a function of the scale $f$ for three values of the parameter
$\kappa$ namely $\kappa = 0.6,1 $ and $1.5$. Although the production cross
section depends only on the mass of the heavy quark, and thus on the 
product $\kappa f$, both the branching 
fractions as well as the decay distributions have additional 
dependence on the scale $f$ and hence we choose to display the three 
curves in Fig.\ref{fig:csprod} so as to facilitate future comparisons.

Once these heavy $T$-odd quarks are produced, they will promptly decay
into ($T$-even) standard model quarks and $T$-odd heavy gauge bosons
$(W^\pm_H, Z_H,A_H)$.  Now, as we have indicated in Sec.\ref{sect:model}, 
the masses of the latter are functions only of $f$. As a comparison of 
eq.(\ref{eq:Toddmass}) with eqs.(\ref{eq:gauge_mass}) shows, $U_H$ and 
$D_H$ are always significantly heavier than the $T$-odd gauge bosons, 
with a slightest hint of phase suppression in $Q_H \to q + Z_H \,  (q' + W_H)$ 
appearing only for the smallest allowed values for $\kappa$. 
More importantly, the 
$Q_H q^{(\prime)} V_H$ couplings too depend on $f$. Whereas the 
couplings $U_H-d-W_H$ and $D_H-u-W_H$ are of equal strength owing to 
$SU(2)$ invariance of the Lagrangian,
\begin{figure}[tb]
\begin{center}
\vspace*{10ex}
\includegraphics[scale=0.45]{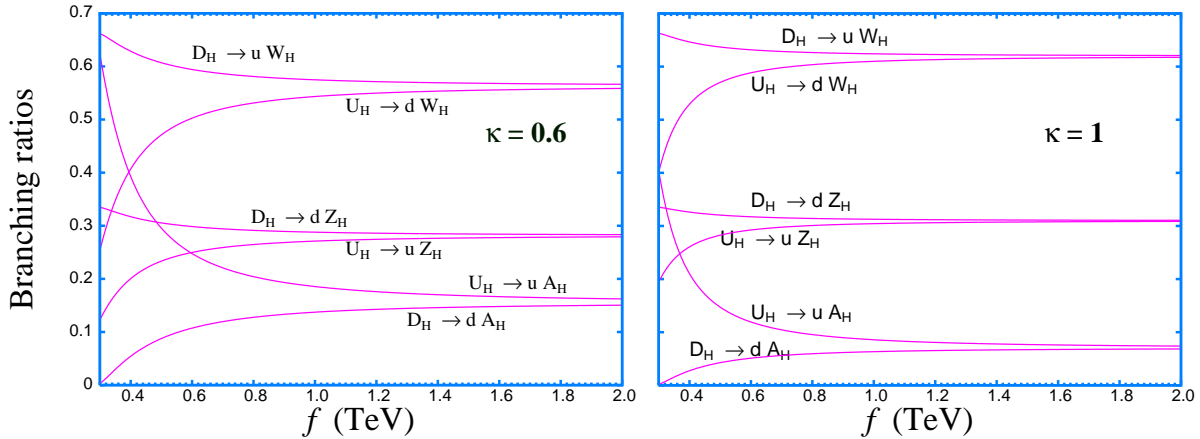}
\end{center}
\vspace*{-9.5cm}
\caption{\em Variation of the decay branching ratio of heavy Quarks in the 
LHT model with the scale $f$ for two values of 
the parameter $\kappa = 0.6$ (left panel) and $1$ (right panel).}
\label{fig:brfig}
\end{figure}
\[
     g_{U_H d W_H} = g_{D_H u W_H} = g / \sqrt{2} \ ,
\]
the couplings to the $Z_H$ and $A_H$ have a crucial dependence 
on isospin ($T_3$), namely
\[
g_{f_H f Z_H} = g \, c_H \, T_{3f} + g' \, s_H \, Y' \ ,
\qquad
g_{f_H f A_H} = - g \, s_H \, T_{3f} + g' \, c_H \, Y' \ ,
\]
where $Y' = - 1/ 10$ and $\theta_H$ is the Weinberg angle in the heavy sector:
\[
  s_H  \equiv \sin\theta_H \simeq
 \frac{5 \, g \, g'} {5 \, g^2 - g'^2} \; \frac{v_{\rm SM}^2}{4 \, f^2}
\ ,
\qquad
  c_H  \equiv \cos\theta_H \ .
\]
This immediately opens up the possibility for a cancellation in 
$g_{D_H d A_H}$ for a relatively small $f$, and consequently 
in the suppression of $\Gamma(D_H \to d + A_H)$ for small $f$. This,
for example, is reflected in 
Fig.~\ref{fig:brfig} where we display the variation of the 
two body decay branching ratios of the $T$-odd quarks into standard
model quarks and heavy $T$-odd gauge bosons as a function of the scale
$f$. 

\section{Signal and background analysis}
\label{sect:sigback}

In this Section, we discuss the LHT signal arising from the production
and decay of heavy $T$-odd quarks of first two generations. We also
discuss possible standard model backgrounds and elaborate on the
selection criteria necessary for such signals to be significantly
observed over the standard model background.
The large number of diagrams contributing to the standard model
background are calculated using the helicity amplitude package {\sc
Madgraph} \cite{madgraph}. To estimate the number of signal and
background events as well as their phase space distribution(s), we use
a parton level Monte-Carlo event generator. As {\em acceptance criteria}
for both the signal and background events we use the following initial set
of cuts:
 \begin{enumerate}
 \item We require that both jets and leptons should appear 
within the detectors' rapidity coverage, namely
\beq
  \mid \eta(\ell, j)\mid <2.5 \ .
     \label{cut:eta}
\eeq 

\item 
The leptons and jets should have energy large enough to render 
them visible to the detector. Imposing this in terms of transverse momenta, 
we demand that
\beq
 p_T^{{\rm jets}} > 30~{\rm GeV} \ , \qquad 
 p_T^{{\ell}} > 20~{\rm GeV} \ .
   \label{cut:pt}
\eeq
\item Finally, we must also ensure that the jets and leptons are well
  separated so that they can be identified as individual entities. For
  this, we use the well-known cone algorithm defined in terms of a 
  cone angle $\Delta R_{\alpha\beta} \equiv \sqrt{ \left
  (\Delta \phi_{\alpha\beta} \right)^2 + \left (\Delta
  \eta_{\alpha\beta} \right )^2} $ with $\Delta \phi$ and $\Delta \eta$ 
  being the azimuthal angular separation and rapidity difference
  between two particles. We demand that 
  \beq
    \Delta R_{jj} > 0.7 \ , \quad
    \Delta R_{\ell j} > 0.4 \ ,\quad
     \Delta R_{\ell \ell}> 0.3 \ .
     \label{cut:deltaR}
   \eeq
 \end{enumerate}
While some of the above might seem too harsh as acceptance criteria,
it should be realized that simulating an actual detector environment
would typically necessitate further refinements and that the requirements
of eqs.(\ref{cut:eta}--\ref{cut:deltaR}) are to be treated more as
robust guidelines. Indeed, harsher requirement on jet rapidity or 
transverse momenta would suppress the QCD background events (wherein
jets come 
from initial or final state radiation) without affecting the signal
to any significant degree.

It stands to reason that finite resolution effects result in a difference 
between the energy as measured by the detector and its true value. To account 
for this in a realistic fashion, we impose a Gaussian smearing on the 
measured energy with a width given by 
\[
\frac{\delta E_j}{E_j} = \left [ \frac{(0.6)^2{\rm GeV}}{E_j} 
+ (0.04)^2\right ]^{1/2} ,
\qquad
\frac{\delta E_\ell}{E_\ell} = \left [ \frac{(0.12)^2{\rm GeV}}{E_\ell} + (0.01)^2\right ]^{1/2} 
\]
respectively.  All the cuts described above as well as any further
selection criteria are to be imposed after smearing the energies as
above.  We may now discuss our strategies for the detection of $T$-odd
heavy quarks at the LHC. For the purpose of contrasting the 
phase space distributions of signal and background events, we choose 
to work with two particular points in the parameter set as displayed in  
Table \ref{tab:param}.
\begin{table}[!h]
\begin{center}
\begin{tabular}{|c|c|c|c|c|c|}
\hline
\multicolumn{6}{|c|}{LHT parameter set}  \\
\hline
\multicolumn{3}{|c|}{\bf A} & \multicolumn{3}{|c|}{\bf B} \\
\hline
\multicolumn{3}{|c|}{$f = 1$~(TeV),~~$\kappa = 0.6 $} &
\multicolumn{3}{|c|}{$f = 1$~(TeV),~~$\kappa = 1.0 $}\\
\hline
$M_{Q_H}~(GeV) $ &  $M_{V_H}$~(GeV) & $ M_{A_H}$~(GeV) & $M_{Q_H}$~(GeV)
& $M_{V_H}$~(GeV) & $ M_{A_H}$~(GeV) \\
\hline
842 & 648 & 154 & 1404 & 648 & 154 \\
\hline
\end{tabular}
\caption{\em The LHT parameter set for the signal study. 
$V_H$ corresponds to $W^\pm_H$ and $Z_H$.}
\label{tab:param}
\end{center}
\end{table}

The simplest final state would arise when both $Q_H$ and $\bar Q_H$
would decay in the $(q + A_H)$ channel. However, the observed final
state, namely dijet with missing transverse momentum is fraught with a
very large SM background. In fact, most final state configurations 
arising as a result of even one of $Q_H$ and $\bar Q_H$ decaying directly 
into  $(q + A_H)$ suffer on this account. In view of such considerations, 
we concentrate on two particular modes as described below.

\subsection{\boldmath $ pp \to Q_H \bar Q_H \to q^\prime {\bar q^\prime}
+ W^+_H W^-_H \to jj + \ell^+ \ell^- + \ptm $} 
This particular final
state arises when both the $T$-odd heavy quarks decay into the ($q
+ W^\pm_H$) mode (with a branching fraction as shown in
Fig.\ref{fig:brfig}). The $T$-odd gauge bosons $(W^\pm_H)$ decay into
the standard model gauge boson $W^\pm$ and the LTP $A_H$ with $\sim 100\%$
branching ratio. And finally, both the $W$'s decay leptonically with
total branching fraction of $\sim (2/9)^2$. The missing transverse energy
$(\ptm )$ is due to the presence of two heavy LTPs $(A_H)$ and two
neutrinos. For ease of detection, we discount $\tau$'s here and hence 
$\ell \equiv e, \mu $. And, while for the signal events 
the jets ($j$) are occasioned by hard processes involving two 
light quarks $(u,d,s,c)$ in the final state, for the SM background one must 
also include hard  gluon(s). 

The major QCD-driven background to this signal emanates 
from the top pair production process
$pp \to t\bar t \to b \bar b W^+ W^- \to b \bar b \ell^-\ell^+ \ptm $
with both $b$-jets being misidentified as light quark jets. Here we
assume that the mis-tagging probability of each $b$-jet as a non-$b$
one is $40\%$. 
The second important source of background 
is the SM process $pp \to W^+W^-jj$, where both $W$s
decay leptonically and the two jets arise from either
quarks or gluons (initial state radiation in the partonic subprocess). 

In addition to $W^+W^-jj$, there are other electroweak processes
contributing to the background, such as $ZZjj$, with one $Z$ decaying
into leptons and the other into neutrinos. Clearly, this background
may be largely eliminated by requiring that the invariant mass of the
lepton-pair be sufficiently away from $M_Z$. In an analogous fashion,
the part of this same background wherein the jet-pair is a resultant
of a $W$ or $Z$ decay, may be further reduced by stipulating that the
dijet invariant mass not be close to either $M_W$ or $M_Z$. In other 
words, our first selection cut (over and above the acceptance 
criteria) consists of
 \beq
M_{jj} \not \in \left[ 65,105 \right ]~{\rm GeV} \; ,\qquad 
M_{\ell \ell} \not \in \left[ 75,105 \right ]~{\rm GeV} \ .
\label{eq:cut_minv_jl}
 \eeq 
Similar arguments also hold for other on-shell modes such as
$W^+ W^- Z$ or $3\, W$'s. Of course, events wherein all the SM gauge
bosons are off-shell escape this cut, but then these appear only at a very
high order in perturbation theory and, consequently, are suppressed.

\begin{figure}[!ht]
\includegraphics[scale=0.8]{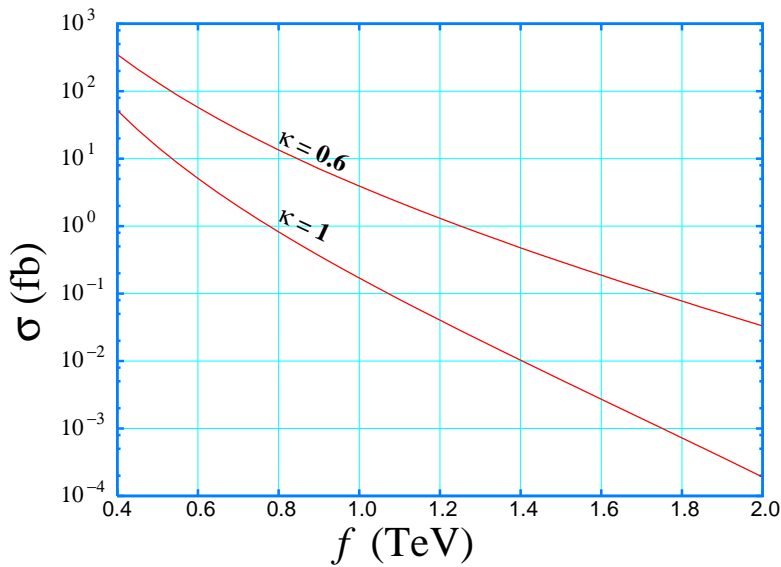}
\caption{\em The variation of the signal $(pp \to Q_H {\bar Q_H} \to 
jj+\ell^+\ell^- + \ptm) $ cross-section with the 
scale $f$ after
imposing the acceptance criteria (eqs.\ref{cut:eta}--\ref{cut:deltaR}) 
as well as the selection cut of eq.(\ref{eq:cut_minv_jl}).}
\label{fig:csjjll}
\end{figure}

\begin{table}[!ht]
\begin{center}
\begin{tabular}{||c||c|c||c|c||}
\hline
Parameter set $\Rightarrow $ & {\bf A} & {\bf B} & \multicolumn{2}{c||}
{$\sigma_{\rm background}$(fb)} \\ \hline
Cuts $\Downarrow$ &$\sigma_{\rm sig.}$(fb)&$\sigma_{\rm sig.}$(fb)&
$\qquad t \bar t \qquad $ & $W^+W^-jj $  \\ \hline 
Acceptance & 4.28~&0.18  & 1095 & 204 \\
\hline
$ M_{jj} \not\in [65, 105]~{\rm GeV}$ & 4.19~&0.18  & 892 &168  \\
\hline
$ M_{\ell\ell} \not\in [75,105]~{\rm GeV} $ & 3.92~&0.17 & 714 &136  \\
\hline
$ \ptm > 200 $ GeV & 2.48~&0.17 &5.6 & 9.33  \\
\hline
$ \ptm > 300 $ GeV & 1.40~&0.13&0.65 &3.12  \\
\hline
$ \ptm > 400 $ GeV & 0.62~&0.10 &0.10 & 1.16  \\
\hline
\end{tabular}
\caption{\em The effect of incremental increase of cuts on the 
signal and background rates (fb) for the process $pp \to Q_H {\bar Q_H}
\to q {\bar q} W^+_H W^-_H \to jj+ \ell^+ \ell^- + \ptm $.
The LHT parameter sets {\bf A} and {\bf B} are defined in 
Table \ref{tab:param}.}
\label{tab:sig_jjll}
\end{center}
\end{table}
Clearly, the signal events are not expected to be affected seriously 
by the imposition of eq.(\ref{eq:cut_minv_jl}), since the jets therein 
arise directly from $Q_H$ decay, whereas the two leptons are the result 
of the decay of two different $W$'s. By the same token, the $t\bar t$
as well as the aforementioned $W^+W^-jj$ background also largely escape
this cut. This is illustrated by Table \ref{tab:sig_jjll}, which displays 
the incremental effect of these two cuts on the major background as well 
as on the signal (for two particular points in the parameter space). 
Of course, the effect of the selection cut (as well as the acceptance 
criteria) on the signal cross section would depend on the masses of the 
$T$-odd quark and gauge bosons, and can be inferred from a comparison of the 
total cross sections (Fig.\ref{fig:csprod}) with the post-cut effective 
cross-sections displayed in Fig.\ref{fig:csjjll}.

\begin{figure}[!ht]
\vspace*{-5.2cm}
\hspace*{-1.5cm}
\includegraphics[scale=0.55]{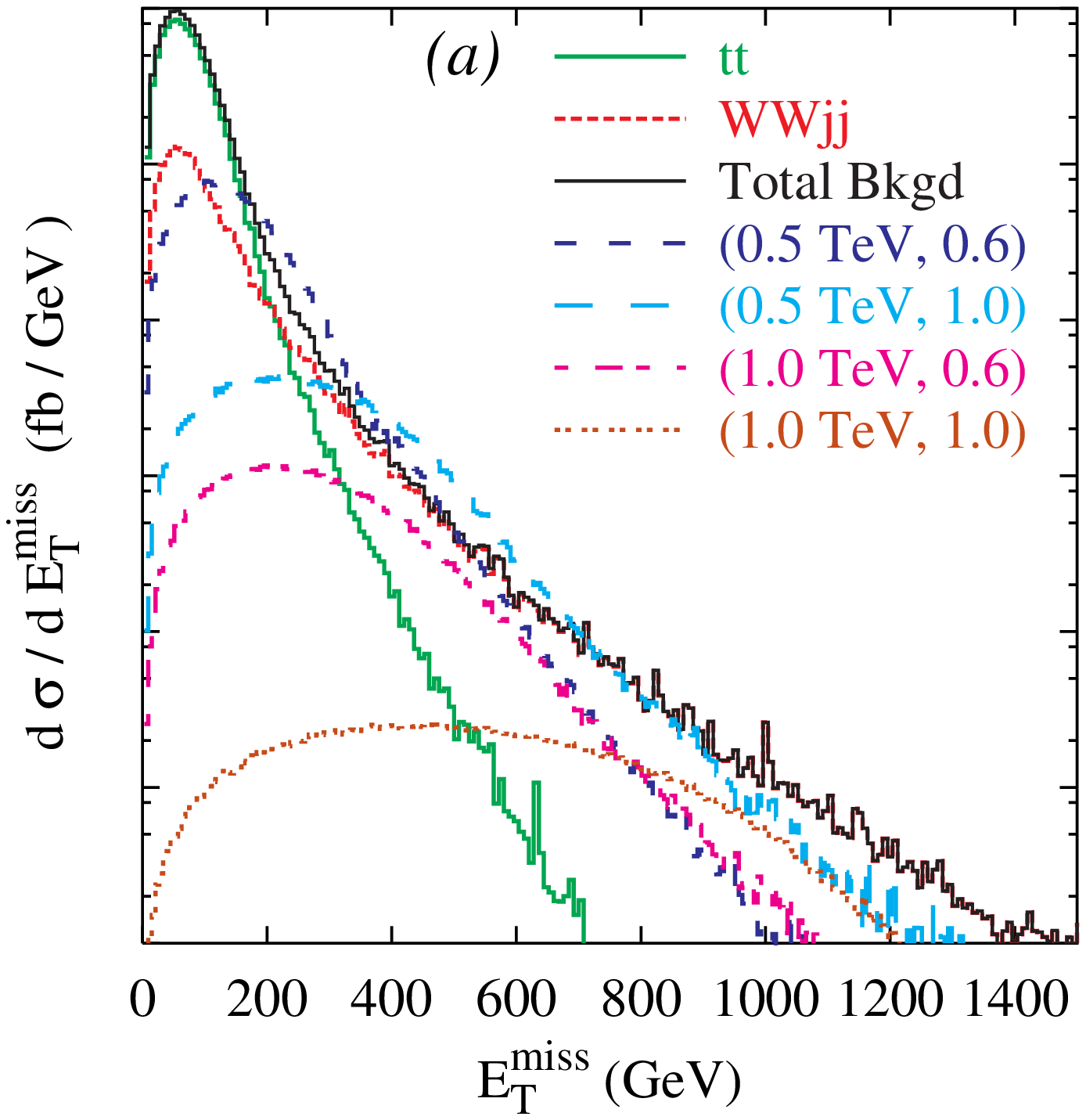}
\hspace*{-2.2cm}
\includegraphics[scale=0.55]{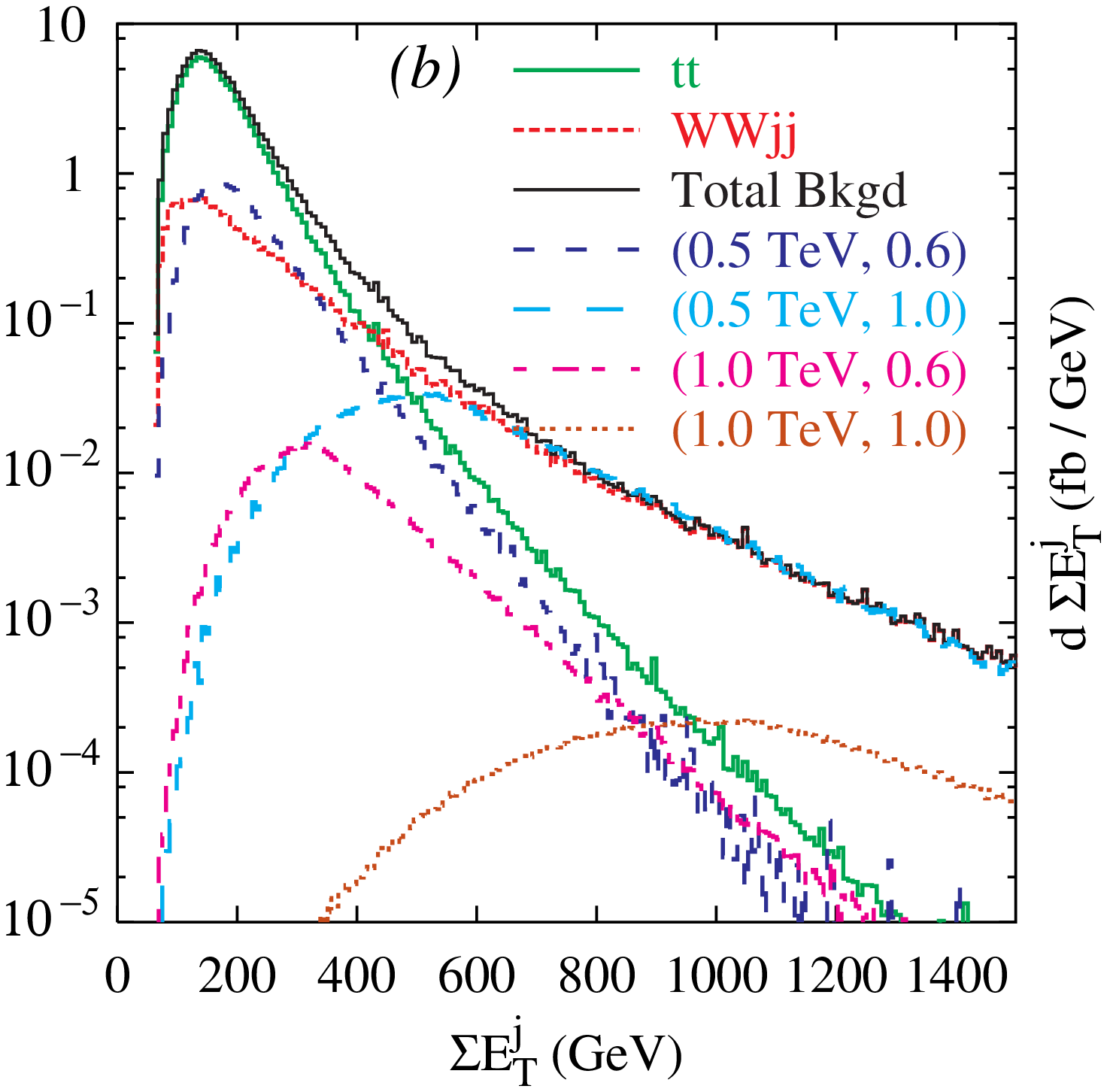}
\vspace*{-1cm}
\caption{\em {\em (a)} Missing $E_T$ distribution 
for the $jj \ell^+\ell^- + \ptm $ final 
state. {\em (b)} Distribution in scalar sum of the jet transverse energies.
Shown are the two dominant SM backgrounds as well as the 
signal for 4 representative points in the $(f, \kappa)$ parameter space.}
\label{fig:dist_ptmjjll}
\end{figure}

As is evinced from Table \ref{tab:sig_jjll}, the number of $t\bar t$
and $W^+W^-jj$ background events which survive eq.(\ref{eq:cut_minv_jl})
are still orders of magnitude higher than the typical signal event
rates. Thus, additional selection criteria are called for. An examination 
of the phase space distributions shows that missing transverse
energy  $(\ptm )$ is a very good discriminatory variable. This is not 
unexpected as the $\ptm$ in the background events arises mainly from 
the two neutrinos, each of which come from the decay of a $W$ and hence 
would typically have a transverse momentum of the order of $m_W$ or smaller.
The signal events, on the other hand, have, apart from the two neutrinos, two 
$A_H$'s each of which are the decay products of a very heavy particle. In 
Fig.\ref{fig:dist_ptmjjll}, we show the differential 
cross sections corresponding to the two major backgrounds as well as 
the signal (4 particular points in the parameter space). 

It is immediately apparent that imposing a strong requirement on
$\ptm$ would result in a significant improvement in the signal to
noise ratio.  In Table \ref{tab:sig_jjll}, we illustrate this for
three choices of $\ptm^{\rm min}$. A second variable of some interest
is the scalar sum of the transverse energies of the two jets. Although 
it is not as discriminatory as $\ptm$, it can be of importance in 
estimating the masses of the quarks and the gauge bosons if a signal
is observed.

\subsection { \boldmath $ pp \to Q_H \bar Q_H \to q^\prime{\bar q}
+ W^\pm_H Z_H \to jj + b \bar b + \ell^\pm + \ptm $}
This final state arises when one of the $T$-odd heavy quarks decays 
into $q + W^\pm_H$ mode, while the other one decays into $q+Z_H$ 
(the third mode, viz. $Q_H \to q + A_H$ can be dominant only for 
small $f$ and that too just for the up-type quarks alone). 
Each of the gauge bosons undergoes a two-body decay to a 
LTP and a SM boson, viz. 
$W_H^\pm  \to W^\pm + A_H$ and
$Z_H \to h + A_H$, with nearly $100\%$ branching ratio. 
And, in the final stages of the cascade, we consider only the 
leptonic decay of the $W$ (branching fraction of $\sim 2/9$), whereas
for the SM Higgs, with an assumed mass of $M_h = 120$ GeV, 
we consider the dominant decay mode, viz. $b \bar b $ (branching 
fraction of $0.68$). 

The collider signature is an interesting one and consists
of an isolated hard lepton $(\ell^\pm)$, four well separated jets 
and a large missing transverse momentum, which owes itself to the 
presence of two heavy LTPs $(A_H)$ and a neutrino from $W$ decay. 
Furthermore, of the four jets, two would be tagged as $b$-jets. 
We assume here that the efficiency for tagging an individual 
$b$-jet is $\epsilon_b = 0.6 $. 

The major background to this particular channel 
comes from the following standard model processes:
\begin{itemize}
\item Top pair production with one top decaying 
hadronically and the other leptonically: 
$pp \to t\bar t \to b \bar b W^+ W^- \to b \bar b 
jj \ell^\pm \ptm $. 
\item $pp \to W^+h jj \to b {\bar b} jj + \ell^\pm \ptm $, 
where the $W$ decays leptonically and $h$ decays
into pair of $b$-jets, while the light quark jets 
originate mainly from initial state radiation.
\item $pp \to W^\pm Z jj \to b {\bar b} jj + \ell^\pm \ptm $, 
where $W$ decays leptonically and $Z$ decays
into pair of $b$-jets. Again, the light quark jets are associated with
initial state radiation.
\end{itemize}

\begin{figure}[tb]
\vspace*{-2cm}
\hspace*{-2cm}
\includegraphics[scale=1.1]{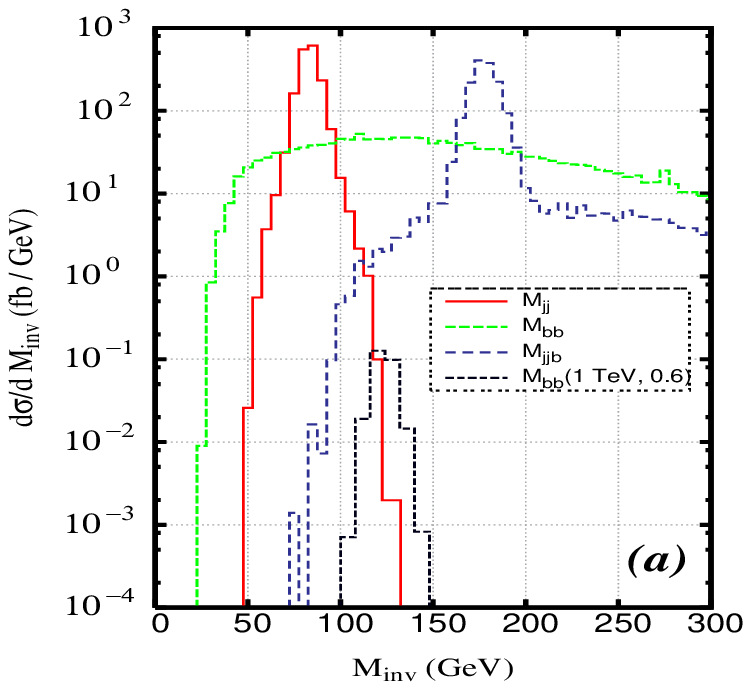}
\hspace*{-1cm}
\includegraphics[scale=0.8]{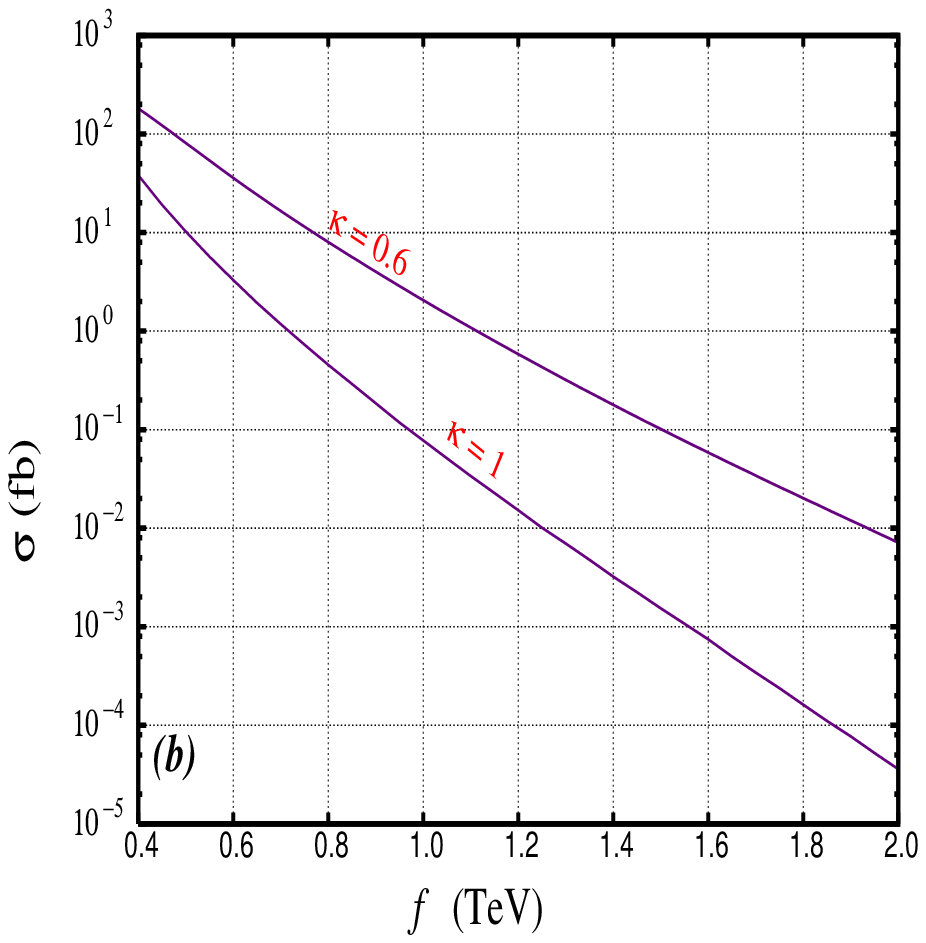}

\caption{\em {\em (a)} $M_{jj},M_{b\bar b}$ and $M_{jjb}$ 
distributions for the standard model background 
to the ($jj+ b{\bar b} +\ell^\pm + \ptm$) final state arising from 
$t\bar t$ production. For comparison, $M_{b \bar b}$ distribution
for the signal ($f=1$ TeV and $\kappa = 0.6$) process is also given. 
 Only the selection cuts (eqs.\ref{cut:eta}--\ref{cut:deltaR}) 
have been applied and the $b$-tagging efficiency included.
{\em (b)} The variation of the signal  cross-section 
with the scale $f$, on imposition of the acceptance cuts 
(eqs.\ref{cut:eta}--\ref{cut:deltaR})
as well as the selection cuts of eqs.(\ref{cut:mjj}--\ref{cut:lnb}). 
}
\label{fig:distminv}
\end{figure}

On imposition of just the acceptance criteria
(eqs.\ref{cut:eta}--\ref{cut:deltaR}), the signal cross-section is
$2.08$~fb and $0.077$ fb for LHT parameter 
sets {\bf A} and {\bf B} respectively, whereas the background arising 
from top pair production is $8930 $ fb as can be seen from 
Table \ref{tab:sig_jjbb}. Clearly, some additional cuts are demanded, 
especially 
to remove the $t \bar t$ background, without suppressing the 
signal cross section. The first such selection criterion is 
exactly the one imposed in the previous subsection, namely that 
the invariant mass of the non-$b$ dijet pair should not be too
close to $M_W$ or $M_Z$. In other words, that 
\beq
    M_{jj} \not \in \left[ 65,105 \right ]~{\rm GeV}.
    \label{cut:mjj}
\eeq
This, clearly, would help eliminate the bulk of the $t \bar t$
background (see Fig.\ref{fig:distminv}). 
In fact, the reduction factor is as large as 100 whereas the 
signal is hardly affected. 
Also eliminated would be the resonant contributions to
the second and third backgrounds listed above, i.e those where the
$jj$ pair resulted from the decay of a gauge boson ($WWh$ and $WZh$
for the second; $WWZ$, $WZZ$ for the third)\footnote{Since these are 
much smaller than the $t \bar t$ background (as well as other 
QCD contributions), we do not list them separately, although we de 
include these in our analysis.}
 
Similarly, since the signal events correspond to the $b$-jets arising from 
the decay of Higgs, we demand that 
\beq
\mid M_{b \bar b}-M_h \mid < 30~{\rm GeV}.
    \label{cut:mhig}
\eeq
The $t \bar t$ background would again be suppressed substantially 
by this requirement as Fig.\ref{fig:distminv} amply suggests. Also suppressed,
to an extent, would be the $WZjj$ background, whereas the $Whjj$ one would 
be largely unaffected. 

Since, for the $t \bar t$ events, the invariant mass $M_{jjb}$ 
constructed from the two untagged jets and one of the $b$-jets would 
cluster around the top mass, we further demand that 
\beq
 \mid M_{jjb}-M_t\mid > 30 \gev \ .
        \label{cut:jjb}
\eeq
for each of the $b$-jets. 
Once again, this requirement would serve to reduce the $t \bar t$ 
background to an extent (see Fig.\ref{fig:distminv}). That this peaking is 
not as sharp as the one for $M_{jj}$ is understandable as this one involves 
measurement of three momenta and hence is subject to larger resolution effects.

At the partonic level, all the missing transverse 
momenta in the $t \bar t$ background events is due to a single neutrino 
born of $W$-decay. Thus, if we equate $ p^T_{\nu} = p^T_{\rm miss}$, the 
longitudinal component of the neutrino momentum 
can be obtained within a quadratic ambiguity 
using the constraint that the invariant mass $M_{\ell \nu} = M_{W}$.
This allows us, then, to reconstruct the second top. 
To further reduce the $t\bar t$ background, we may then demand 
that the invariant mass of the ($\ell\nu b$) combinations 
should not match $M_t$:
\beq
 \mid M_{\ell \nu b}-M_t\mid > 30 \gev \ .
        \label{cut:lnb}
\eeq
\begin{table}[!h]
\footnotesize
\begin{center}
\begin{tabular}{|c|c|c|c|c|c|}
\hline
Parameter set $\Rightarrow $ & {\bf A} & {\bf B} & \multicolumn{3}{|c|}
{$\sigma_{\rm background}$(fb) } \\ \hline
Cuts $\Downarrow $ & $\sigma_{\rm sig.}$(fb) & $\sigma_{\rm sig.}$(fb) &
$t \bar t $ & $W^\pm h jj $ & $ W^\pm Z jj $ \\ 
\hline
Acceptance & 2.08 &0.077 &8930 & 12 & 35.54  \\
$ M_{jj} \not\in [65, 105]~{\rm GeV}$   & 2.04 &0.077 &88.36 & 10.1 & 30.02 \\
$ \mid M_{b \bar b}-M_h \mid < 30~{\rm GeV} $ 
& 2.04 &0.077 &27.29 & 9.45  &18.65  \\
$ \mid M_{jj b}-M_t \mid \ , \mid M_{\ell\nu b}-M_t \mid > 30~{\rm GeV} $ 
 & 2.03 &0.077 &1.26 &  9.41 & 18.57 \\
$ \ptm > 200 $ GeV & 1.41 &0.069 &$\sim {\cal O}(10^{-4})$& 0.21   & 0.47 \\
$ \ptm > 300 $ GeV &0.84  &0.06&$ \lsim {\cal O}(10^{-5})$ &0.043 & 0.11 \\
$ \ptm > 400 $ GeV & 0.40 &0.05 &$\lsim  {\cal O}(10^{-7})$ &0.010 & 0.038\\
\hline
\end{tabular}
\caption{\em The incremental effect of cuts on the 
signal and background rates for the process $pp \to Q_H {\bar Q_H}
\to q {\bar q} W^\pm_H Z_H \to b {\bar b} j j+ \ell^\pm + \ptm $.
The LHT parameters are as in Table \ref{tab:sig_jjll}.
The $b$-tagging efficiency has been included.}
\label{tab:sig_jjbb}
\end{center}
\end{table}

As Table~\ref{tab:sig_jjbb} shows, the imposition of the selection
criteria of eqs.(\ref{cut:mjj}--\ref{cut:lnb}) results in suppressing
the $t \bar t$ background by a factor $\gsim 7000$ while leaving the
signal size essentially unaltered. Also reduced significantly is the
$W^\pm Z jj$ background, whereas the $W^\pm h jj$ suffers only a minor
reduction. However, owing to their large initial sizes, they still
dominate the signal over the entire LHT parameter space. Indeed, as even a
cursory comparison of Figs.\ref{fig:distminv} shows, for $m_{Q_H}
\lsim 1400 \gev$, the sensitivity, at this stage, is
background-limited rather than signal-limited. This, then, motivates
the introduction of further selection cuts, and once again we consider
the missing transverse momentum as well as $\sum E_T^j$, the scalar
sum of the transverse energies of the two non-$b$ jets. 

\begin{figure}[tb]
\vspace*{-5.2cm}
\hspace*{-1.5cm}
\includegraphics[scale=0.55]{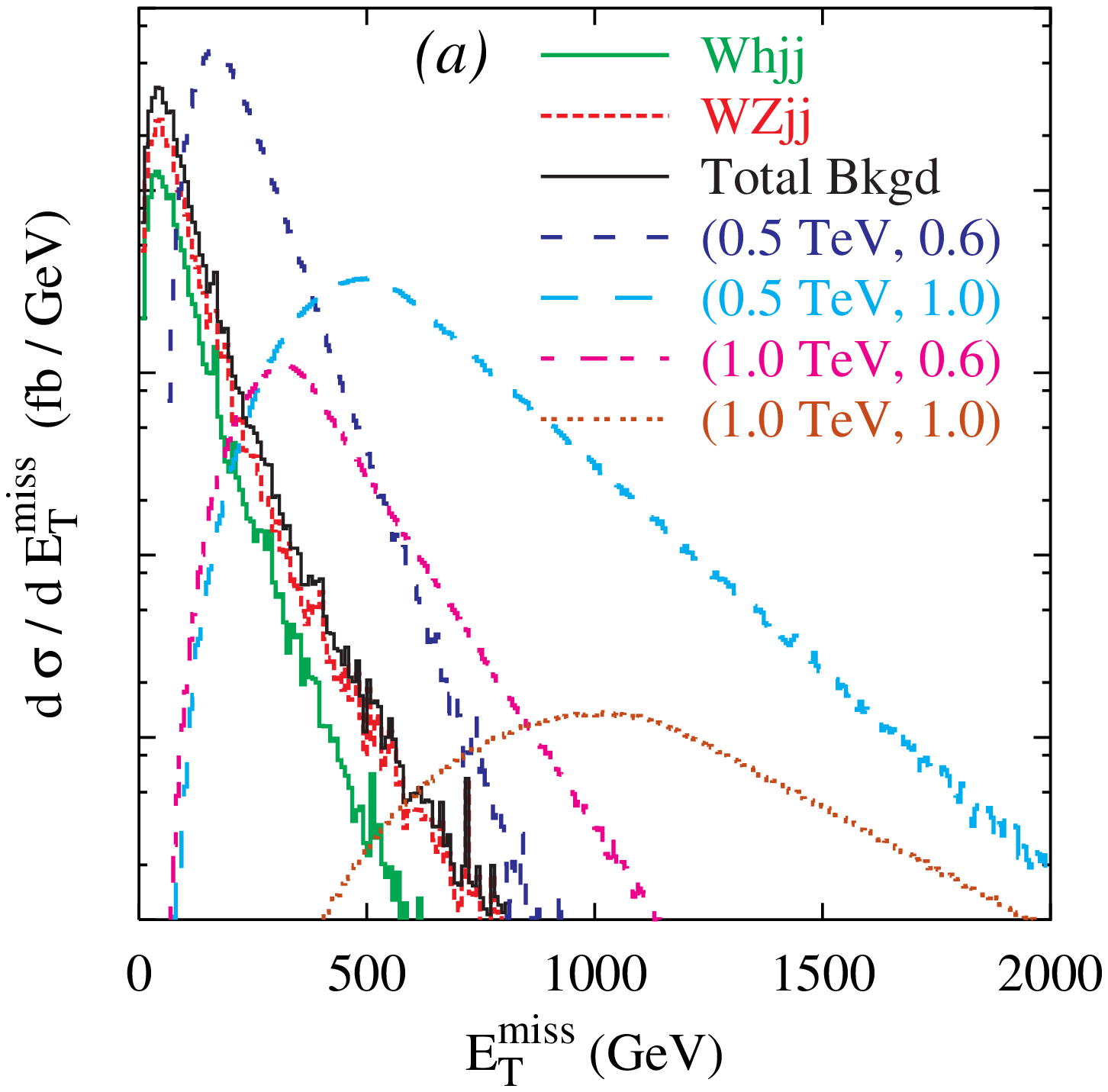}
\hspace*{-2.2cm}
\includegraphics[scale=0.55]{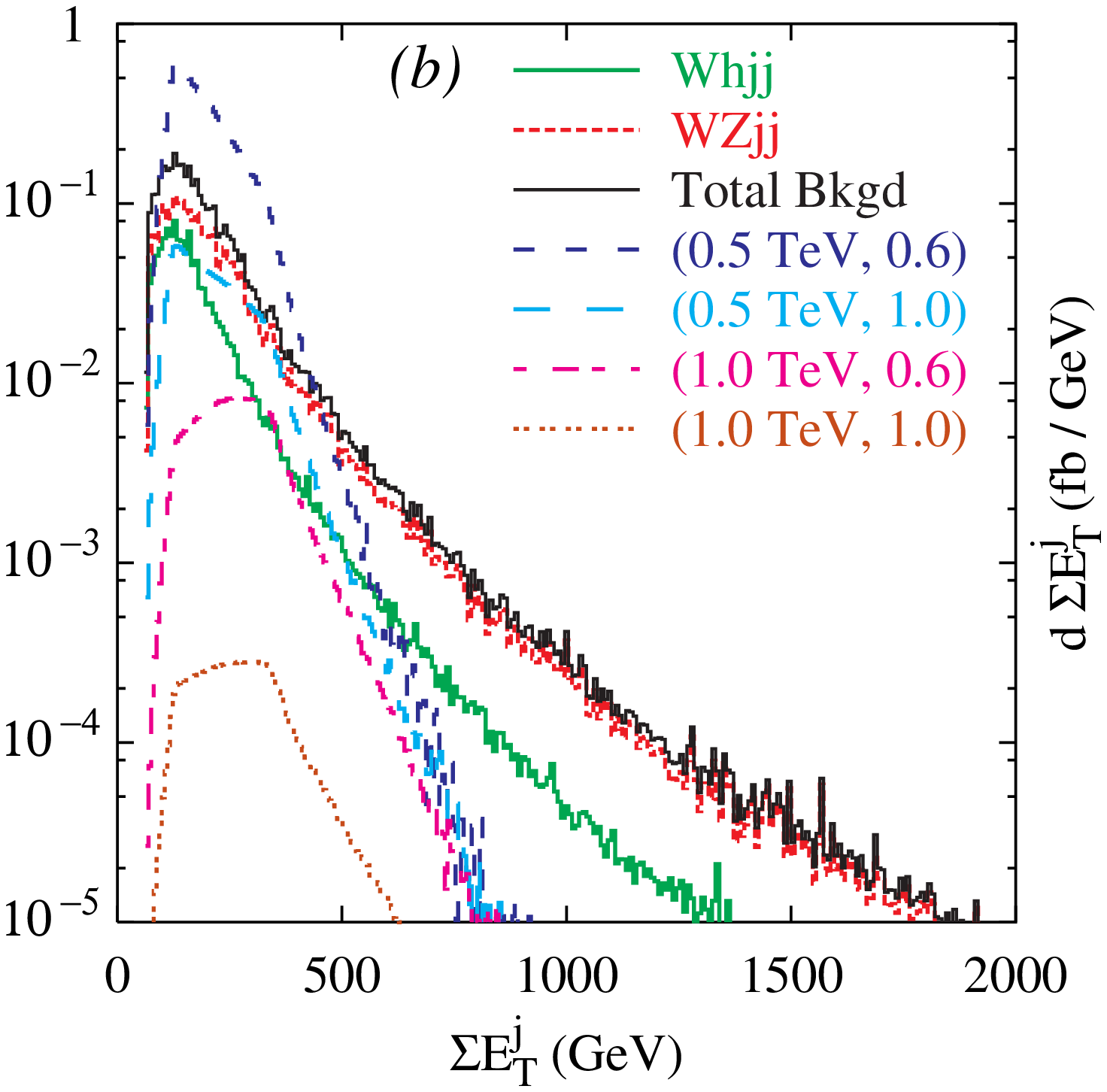}
\vspace*{-1cm}
\caption{\em {\em (a)} Missing $E_T$ distribution 
for the $jj b \bar b \ell^\pm + \ptm $ final 
state. {\em (b)} Distribution in scalar sum of the two non-$b$ 
jet transverse energies.
Shown are the two dominant SM backgrounds as well as the 
signal for 4 representative points in the $(f, \kappa)$ parameter space.}
\label{fig:dist_ptmbb}
\end{figure}

As Fig.\ref{fig:dist_ptmbb} shows, the background $\ptm$ distribution
is much softer in this case (as compared to that in
Fig.\ref{fig:dist_ptmjjll} for the signal considered previously). This
is understandable as the final state now has only one neutrino rather
than the two for the previous case. And while the corresponding
distributions for the signal events are softer too (again due the
decrease in the number of neutrinos), the reduction is not
severe. This, in part, is due to the fact that a large part of $\ptm$
accrues on account of the the two (heavy) $A_H$'s. The difference in the
small $\ptm$ end of the spectrum is attributable to the fact that, for
the ($ jj + \ell^+ \ell^- +\ptm $) case, the requirement on a minimum
transverse momenta for both the leptons generically implies a larger
$\ptm$ as well. In all, thus, the imposition of an identical cut on
$\ptm$ serves to improve the signal to background ratio for the ($jj +
b \bar b + \ell^\pm + \ptm $) signal to a much larger degree than was
the case for the ($ jj + \ell^+ \ell^- +\ptm $) one. The quantitative
effect can be gauged by a perusal of Table~\ref{tab:sig_jjbb}. Of particular 
interest is the fact that the ordinarily dominating $t \bar t$ can be 
eliminated to the extent of less than one event satisfying the selection 
criteria during the entire planned run of the LHC. 

While the distribution in $\sum E_T^j$ continues to be less
discriminatory than the one in $\ptm$ (see Fig.\ref{fig:dist_ptmbb}),
an examination of the same is, nevertheless, quite
instructive. Naively, for the signal events, one would have expected
this distribution to look very similar for the ($ jj + \ell^+ \ell^-
+\ptm $) and ($jj + b \bar b + \ell^\pm + \ptm $) cases, since the
jets are occasioned in both cases by the decay of the $Q_H$ to a SM
quark and a $W_H$ or $Z_H$ (with the bosons being very close in
mass). That the spectra look a little different is
attributable to the effect of the kinematical cuts which, of course,
are different in the two cases. Once again, the distribution for the
background is softer in the present case as compared to the previous
one. As Fig.\ref{fig:dist_ptmbb} suggests, it would be profitable to
exploit a combination of cuts on $\ptm$ and $\sum E_T^j$, so as to
improve the signal to background ratio, but given the rather sharp
improvement from a consideration of $\ptm$ alone, we desist from doing
this.

\section{Discovery Limit}
\label{sect:discover}
Having established that a suitable choice of selection criteria can
serve to suppress the admittedly large SM background, thereby
enhancing the signal profile (for at least some parameter choices
studied above), we now examine the extent to which this can be done.
As a comparative study of Fig.\ref{fig:dist_ptmjjll} and
Fig.\ref{fig:dist_ptmbb} immediately shows, the ($jj + b \bar b +
\ell^\pm + \ptm$) final state is expected to have a far better signal
to noise ratio than the ($jj + \ell^+ \ell^- + \ptm $) one. We may thus 
safely concentrate on the former in our efforts to delineate the 
parameter space.

\begin{figure}[!h]
\vspace*{-4.0cm}
\hspace*{-1.5cm}
\includegraphics[scale=0.62]{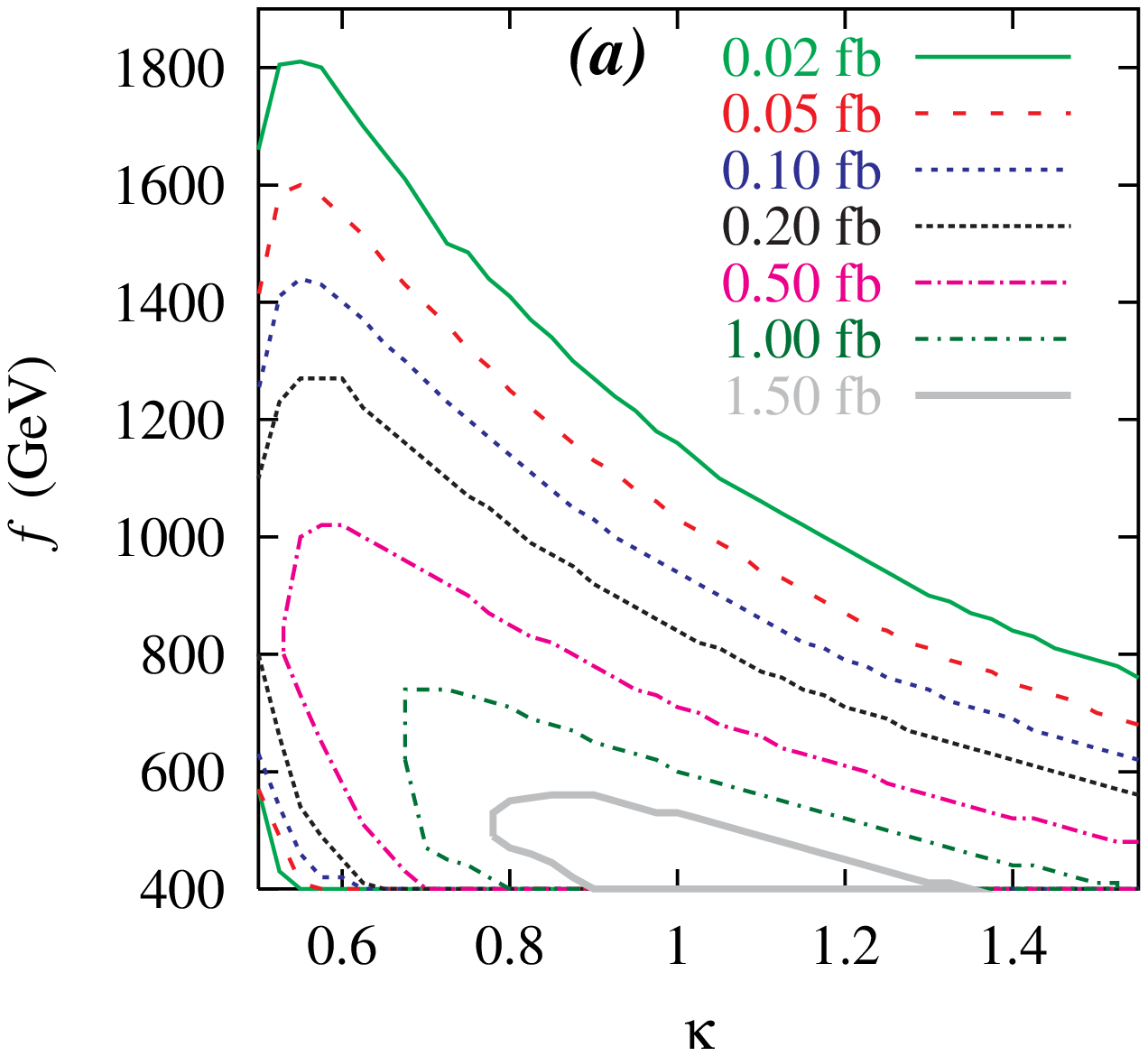}
\hspace*{-0.8cm}
\includegraphics[scale=0.62]{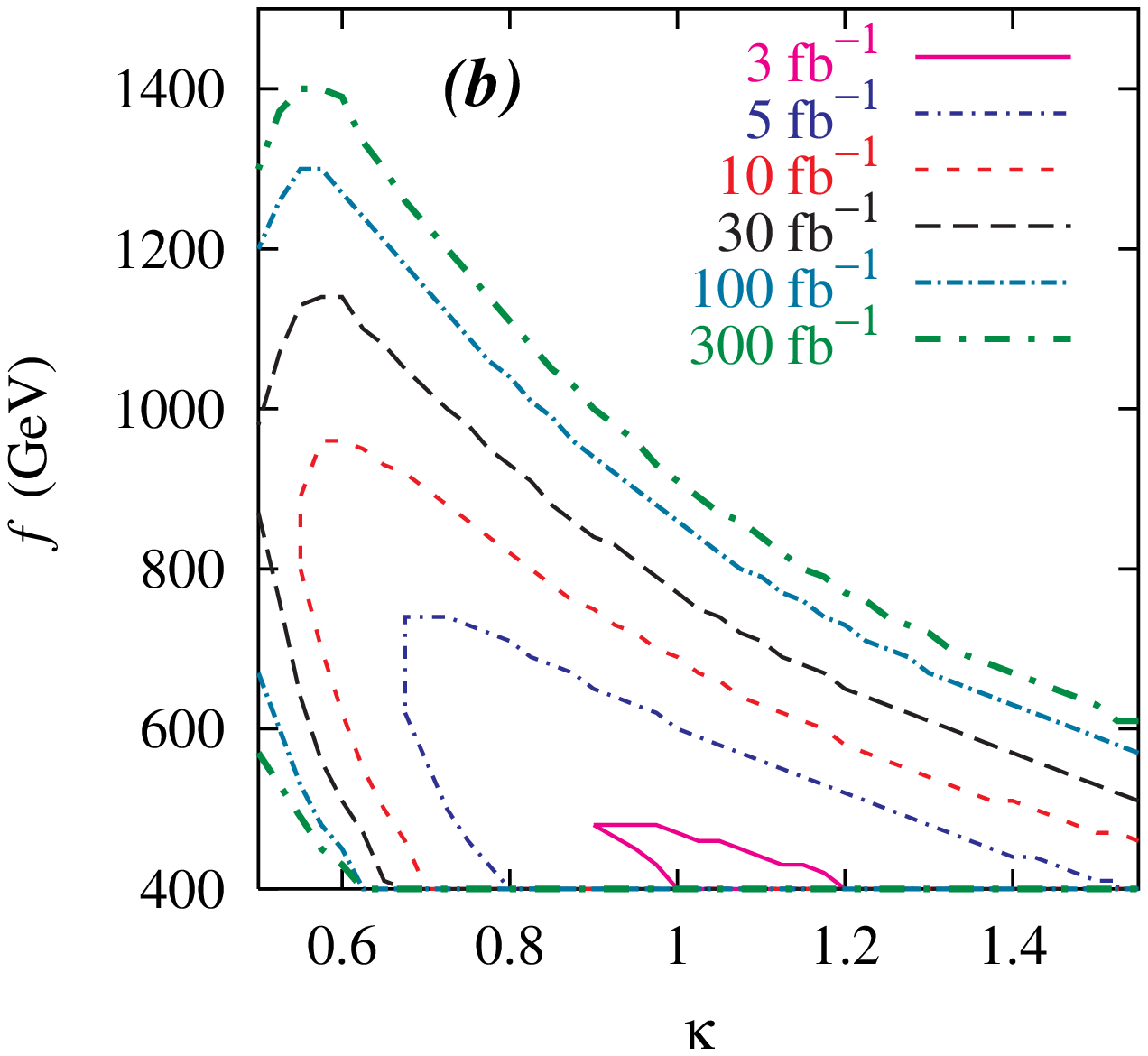}

\vspace*{-1.2cm}
\caption{\em {\em (a)} Constant cross section contours in the 
$\kappa$-$f$ plane for the ($jj+ b{\bar b} +\ell^\pm + \ptm$) final state.
Apart from the acceptance cuts (eqs.\ref{cut:eta}--\ref{cut:deltaR}), 
thee selection cuts of eqs.(\ref{cut:mjj}--\ref{cut:lnb}) and a further 
requirement of $\ptm > 400 \gev$ have been imposed. 
{\em (b)} The associated $5 \sigma \, (1 - {\rm C.L.} = 
5.7 \times 10^{-7})$ contours for different values of the 
integrated luminosity.
}
\label{fig:contours}
\end{figure}

In Fig.\ref{fig:contours}($a$), we present constant cross section
contours for the same. Since the requirement of $\ptm > 400 \gev$
eliminates virtually all of the background (vide
Table~\ref{tab:sig_jjbb}), we have chosen to impose this. As is
expected, for much of the parameter space, the cross section is
primarily a function of the combination ($\kappa \, f$) alone. At low
$\kappa$ and low $f$ though, the smallness of $A_H$ mass results in a
suppression of the total missing transverse energy and hence to a
relatively larger loss due to the cut on $\ptm$.  Similarly, the
smallness of the masses of the other $T$-odd particles ($Q_H, W^\pm_H,
Z_H$) results in the daughter particles having smaller energies leading to
a loss on account of the other selection cuts. 

This, then, reinforces the argument of the previous section in favour
of either mass-dependent selection cuts or the comparison of
multivariate event distributions for both signal and background (a la
unbinned likelihood analysis). However, bearing in mind the nature of
this analysis, we deliberately choose not to adopt such sophisticated
tools and restrict ourselves to just the set of mass-dependent
selection cuts mentioned above. This, of course, amounts to a conservative
choice. Since both the signal and background
events are small in number, 
we estimate the discovery limit in the LHT parameter
space, assuming that they follow the
well known Poisson distribution. Thus, a $5\sigma $ discovery corresponds
to $1- \alpha \le 5.7\times 10^{-7}$, with $\alpha (N_0)$ 
being the Poisson probability for seeing upto $N_0$ events
when $N_b$ background events are expected.  In
Fig. \ref{fig:contours} ($b$) we show the $5\sigma $ discovery region
in the LHT parameter space by using the signal topology of $jj + b
\bar b + \ell^\pm + \ptm $.  As Fig.\ref{fig:contours}($b$) amply
exhibits, even with a single year of low-luminosity run $(L = 10~{\rm
fb}^{-1})$, a remarkable part of the LHT parameter space can be
probed. For the highest luminosity, the reach can be
further improved, with $f$ being probed all the way upto 1.4 TeV for
$\kappa = 0.6$, while $\kappa $ can be probed upto 1.5 for $f \sim 600
$ GeV. Conversely, for optimistic values of the parameters, a discovery can be
made with only a few months running time.

\section{Conclusions}
\label{sect:concls}
In this paper, we have discussed two types of signatures of the first 
two generations of heavy $T$-odd quarks predicted by the Littlest Higgs model
(LHT). It has been shown that $T$-odd heavy
quarks can be copiously pair produced ($Q_H \bar Q_H$) at the LHC 
as long as their masses are not too large \cite{belyaev, wyler,
carena}. As the heavy quarks corresponding to the first two generations are
nearly degenerate (Sec. \ref{sect:t-oddprod}), and lead to very
similar final state configurations, we summed over all four flavours.  In
our numerical analysis, we have used the CTEQ5L parton distribution
functions\cite{cteq5}. Whereas the production cross
section depends only on the mass of the heavy quark, and hence on the 
product $\kappa f$, both the branching 
fractions as well as the decay distributions have additional 
dependence on the scale $f$ as we have discussed in Sections \ref{sect:model}
and \ref{sect:t-oddprod}. 
Once these heavy $T$-odd quarks are produced they will promptly decay
into ($T$-even) standard model quarks and $T$-odd heavy gauge bosons
$(W^\pm_H, Z_H,A_H)$ with appropriate branching ratios which depends 
upon the scale $f$ and $\kappa $ as we have shown in Fig. \ref{fig:brfig}.

We mainly focussed on the following two types of signal configurations, viz.
$(a)~pp \to Q_H \bar Q_H \to q^\prime {\bar q^\prime}
+ W^+_H W^-_H \to jj + \ell^+ \ell^- + \ptm $ and 
$(b)~pp \to Q_H \bar Q_H \to q^\prime{\bar q}
+ W^\pm_H Z_H \to jj + b \bar b + \ell^\pm + \ptm $. 
The major background for the signal
type $(a)$ comes from the standard model processes
$t \bar t$ and $W^+W^-jj $, whereas the standard model processes 
$t\bar t$, $W^+hjj$ and $W^\pm Z jj$ comprise the major 
backgrounds for the signal type $(b)$. 
To estimate the number of signal and
background events as well as their phase space distribution(s), we have 
used a parton level Monte-Carlo event generator. At first, we forced 
both signal as well as background events to satisfy acceptance criteria
as discussed in Section \ref{sect:sigback}. We have then selected two sets 
of LHT parameters as displayed in 
Table~\ref{tab:param} for the purpose of comparing differential distributions 
as well as total cross-sections of signal and background events. 
It was found that 
the standard model background rates were order of 
magnitude higher than that of the signal events even after satisfying 
our acceptance and preliminary selection cuts. Hence, additional set of 
selection cuts were required 
to improve the signal rates. After studying distributions of different 
kinematic variables, we find that the missing transverse energy $(\ptm)$
would provide a good discriminator. 
As Fig.\ref{fig:dist_ptmjjll} shows, even after a stringent cut on 
$\ptm > 400 $ GeV, signal $(a)$ can supersede the background only for a
small range of LHT parameters.
However, for signal $(b)$, we find a rather encouraging
situation, as all three standard model background rates turn out to be 
significantly smaller than the signal rates once we impose the 
cut $\ptm > 400 $ GeV as shown in the Table \ref{tab:sig_jjbb}. Consequently,
$pp \to Q_H \bar Q_H \to q^\prime{\bar q}
+ W^\pm_H Z_H \to jj + b \bar b + \ell^\pm + \ptm $ constitutes the dominant
discovery channel for the first two generation $T$-odd heavy quarks at the
LHC. Using this particular channel we have obtained 
$5\sigma $ discovery limit in the LHT parameter space.  
As Fig.\ref{fig:contours}($b$) amply shows, 
adopting this methodology would allow us to make a discovery 
over a significant area in the allowed parameter space with only a few 
months' worth of data. For higher luminosities, the 
LHT scale $f$ can be probed all the way upto $\sim {\cal O}~({\rm TeV})$ using
this $jj + b \bar b +\ell^\pm+\ptm $ channel.  
We, thus, expect that the parton level study presented in this paper will 
encourage
the CMS and ATLAS collaboration to carry out further investigations
of the Littlest Higgs Model with $T$-parity.

\noindent 
\section*{Acknowledgments}
The authors acknowledge several useful discussions with Satyaki 
Bhattacharya and Sukanta Dutta. DC acknowledges support from the 
Department of Science and Technology, India under project number:
SR/S2/RFHEP-05/2006.

\end{document}